\documentclass[%
reprint,
superscriptaddress,
floatfix,
amsmath,
amssymb,
aps,
notitlepage
]{revtex4-2}
\usepackage{mathrsfs,stmaryrd}
\usepackage{amsmath}
\usepackage{graphicx,setspace}
\usepackage{bm}
\usepackage[hidelinks=true]{hyperref}
\usepackage{xcolor}
\definecolor{greenish}{RGB}{41,198,63}
\definecolor{reddish}{RGB}{239,65,54}
\definecolor{blueish}{RGB}{28,117,188}
\definecolor{magenta}{rgb}{0.8, 0.0, 0.8}
\hypersetup{
colorlinks   = true, 
urlcolor     = greenish, 
linkcolor    = reddish, 
citecolor   = blueish 
}

\def\u{\mathbf{u}}

\def\Ca{\mathsf{C}_a}

\def\iota{i}

\let\origmathbf\mathbf
\renewcommand{\mathbf}[1]{\ifmmode\origmathbf{#1}\else\textbf{#1}\fi}

\newcommand{\bu}{\boldsymbol{u}}

\begin{document}

\title[An \textsf{achemso} demo]
  {Morphological instability of an invasive active-passive interface}

\author{Sumit Sinha}
\affiliation{School of Engineering and Applied Sciences, Harvard University, Cambridge MA 02138.}
\author{Haiqian Yang}
\affiliation{Department of Mechanical Engineering, Massachusetts Institute of Technology, 77 Massachusetts Ave., Cambridge, MA 02139, USA}
\author{L. Mahadevan}
\email{lmahadev@g.harvard.edu*}
\affiliation{School of Engineering and Applied Sciences, Harvard University, Cambridge MA 02138.}
\affiliation{Department of Physics, Harvard University, Cambridge MA 02138.}
\affiliation{Department of Organismic and Evolutionary Biology, Harvard University, Cambridge 02138}


\begin{abstract}
\textcolor{black}{Morphological instabilities of growing tissues that impinge on passive materials are typical of invasive cancers. To explain these instabilities in experiments on breast epithelial spheroids in an extracellular matrix, we develop a continuum phase field model of a growing active liquid expanding into a passive viscoelastic matrix. Linear stability analysis of the sharp-interface limit of the governing equations predicts that the tissue interface can develops long-wavelength instabilities, but these instabilities are suppressed when the active carcinoid is embedded in an elastic matrix. We develop a theoretical morphological phase diagram, and complement these with two-dimensional finite element (FEM) phase-field simulations to track the nonlinear evolution of the interface with results consistent with theoretical predictions and experimental observations. Our study provides a basis for the emergence of interfacial instabilities in active-passive systems with the potential to control them.}
\end{abstract}

\date{\today}
\maketitle

\noindent \textit{Introduction:}
The invasive nature of tumors and their ability to metastasize to distant organs remain among the primary causes of cancer-related mortality. A central mechanism underlying this process is collective cell migration, through which groups of tumor cells coordinately invade surrounding tissue \cite{friedl2012classifying, cheung2016polyclonal}. This mode of invasion is characterized by strong mechanical coupling, sustained tissue-level deformation, and emergent collective dynamics.

A variety of experimental and theoretical approaches have been used to study collective tumor invasion. Experimental studies have focused on identifying histological, biochemical, and phenotypic signatures associated with collective migration \cite{lee2021distinct}. Complementing these, agent-based and particle-based models have been developed to resolve cell-scale mechanics, adhesion, and rearrangements during invasion \cite{malmi2018cell, drasdo2005single, matoz2017cell}. At larger scales, continuum descriptions have treated tissues as active materials, employing hydrodynamic theories of active polar and nematic fluids \cite{Ranft10PNAS, bogdan2018fingering, alert2019active, alert2022fingering}, chemotactic fronts \cite{alert2022cellular}, and related formulations to study interfacial instabilities, tissue fluidization, and fingering phenomena \cite{doostmohammadi2015celebrating, bogdan2018fingering, ranft2010fluidization, monfared2025multiphase}. These continuum approaches have also been extended to include environmental effects through mechanically imposed boundary stresses \cite{Basan11PRL, williamson2018stability}.

\begin{figure}
    \includegraphics[width=1.1\columnwidth]{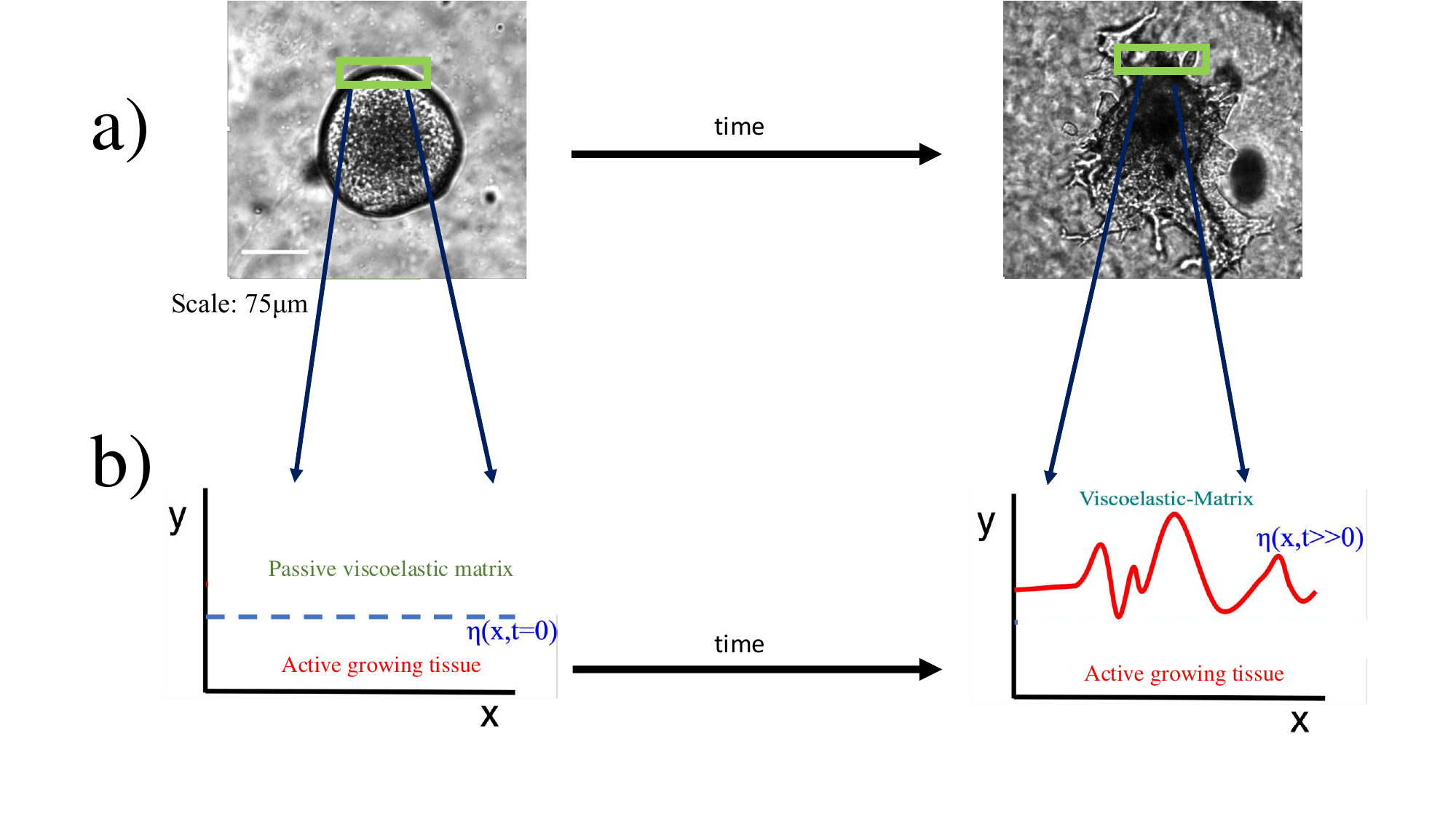}
    \caption{\textbf{Viscoelasticity of the extra-cellular matrix regulates the morphological instabilities at an invasive active-passive interface.} (a) The top row shows the development of instabilities at the interface when a an active growing spheroid is placed in a liquid like matrix ($\tau\rightarrow 0$). Figure taken from the study of Elosegui-Artola et.al.\cite{elosegui2023matrix}. (b) Zooming in allows us to focus on the interface. In the sharp-interface limit, we assume that an active growing tissue moves and pushes against a viscoelastic (linear Maxwell model) matrix. The  interface, $\eta(x,t)$, is  initially flat at $t=0$ (dashed blue line) but can develop instabilites at a later time. In a minimal setting, we assume that the tissue-matrix lies in the $x-y$ plane, along the x-axis extending from $-\infty \leq x \leq \infty$.}
    \label{fig1}
\end{figure}

Recent experiments \cite{elosegui2023matrix} have added a new dimension to this picture by demonstrating that the viscoelastic properties of the extracellular matrix play a decisive role in regulating tumor invasion. In particular, the interface of an actively growing breast epithelial tumor spheroid remains flat or spherical when embedded in an elastic matrix, but becomes unstable to long-wavelength undulations when the matrix is more liquid-like. This behavior can be understood by comparing the matrix stress-relaxation timescale with the tissue growth timescale: if the matrix stress relaxes slowly, this suppresses interfacial instabilities, whereas if the matrix relaxes quickly, there is a propensity for enhancing instabilities. These experimental observations were rationalized using agent-based cellular simulations that explicitly model cell–matrix interactions and matrix remodeling \cite{elosegui2023matrix}. While these simulations successfully reproduce the observed phenomenology, they do not yet provide a general theoretical framework that links matrix viscoelasticity, active growth, and interfacial stability in a transparent and predictive manner.

Here, we develop a continuum phase-field hydrodynamic theory, together with finite-element simulations, to provide such a framework. We build on phase-field methods for modeling interfacial dynamics in complex fluids and soft matter \cite{anderson1998diffuse, provatas2010theory,lowengrub2009phase}, by  coupling interfacial motion to hydrodynamics and active stresses in growing tissues interacting with viscoelastic environments to explain the experimentally observed behavior of the spheroid–matrix interface in both the linear and nonlinear regimes.

\noindent \textit{Diffuse-interface model of an active tissue invading a viscoelastic matrix:} We model a growing epithelial spheroid invading a viscoelastic extracellular matrix using a diffuse-interface (phase-field) formulation that unifies growth, active forcing, matrix viscoelasticity, and interfacial capillarity within a single continuum framework.  In  Fig.\ref{fig1}a, we show the evolution of the instability of a growing epithelial spheroid invading the extracellular  matrix, and note that the basic spatiotemporal fields that need to be tracked include the active tissue phase and passive viscoelastic matrix, characterized by a velocity field ${\bf u(r},t)$, and the stress-field $\boldsymbol{\sigma}({\bf r},t)$. The diffuse-interface (phase-field) is modeled via the scalar order parameter $\phi(\mathbf{x},t) \in [-1,1] $ which distinguishes the active tissue ($\phi=+1$) from the passive matrix ($\phi=-1$), with a narrow interfacial region of thickness $\epsilon$, assumed to be smaller than all other length scales.  In addition to resolving the microscopic physics in a thermodynamically consistent manner, the phase-field model is much easier to solve numerically without the need to separately track the interfaces.  

The phase-field, $\phi$, evolves according to a Cahn-Hilliard like equation, 
\begin{equation}
\frac{\partial \phi}{\partial t} + \mathbf{u} \cdot \nabla \phi 
 -M \nabla^2 c - \mathcal{G}(\phi)=0, 
 \label{eqn:pfch}
\end{equation}
modified to account for the advection of the phase(s) by the fluid with velocity $\mathbf{u}({\bf x},t)$, and a growth term given by $\mathcal{G}(\phi)=\gamma\exp(-\phi^2)(1-\phi^2)=\gamma g(\phi)$. The form of the growth has been chosen so that growth near the interface (i.e. $\phi \approx 0$) is promoted. Here the double-welled free energy density ${\mathcal F}/\kappa_{\rm ch} = (1-\phi^2)^2+ \epsilon^2 (\nabla \phi)^2$ , so that the chemical potential$ c =\frac{\delta \mathcal F}{\delta \phi}= \kappa_{\rm ch}(\phi^3 - \phi - \epsilon^2\nabla^2\phi)$, with $\kappa_{\rm ch}$ having dimensions of energy density, and $M$ characterizing the mobility.

Force balance (in the overdamped, low Reynolds number limit) for the multi-phase mixture reads as
\begin{equation}
\nabla \cdot \big(\boldsymbol\sigma_N + f(\tau)\frac{(1-\phi)}{2}\boldsymbol\sigma' \big) 
+ c \nabla \phi
+ \alpha(\phi){\bf u} = 0, 
\label{eqn:force_bal}
\end{equation}
Here the total stress is assumed to be the sum of the phase-dependent Newtonian stress $\boldsymbol\sigma_N$ and a viscoelastic contribution for the matrix denoted by $\boldsymbol\sigma'$. The additional terms in (2) are the diffuse capillary (Korteweg) force $c\nabla \phi$ that reduces to the Laplace pressure jump in the sharp-interface limit \cite{zhang2010simulating,yue2006phase}, and an activity due to the movement of cells in the growing tissue that we assume is proportional to $\alpha(\phi)\u$, where $\alpha(\phi)$ is the phase-dependent part assumed to be of the form $\alpha(\phi)=\tfrac{1+\phi}{2}\alpha_t$ which vanishes in the passive tissue (when $\phi \approx -1$.

The Newtonian stress is $\boldsymbol\sigma_N=-p {\mathbf{I}} + 2\eta_s(\phi)D({\bf u})$, where $p$ is the pressure, and the strain rate $D({\bf u})=\frac{1}{2}(\nabla {\bf u} + \nabla {\bf u}^T)$ with  a simple linear interpolant for the viscosity of the mixture of the form $\eta_s(\phi)=\frac{(1+\phi)}{2}\eta_{s,t}+\frac{(1-\phi)}{2}\eta_{s,m}$, where $\eta_{s, t}$ and $\eta_{m, t}$ are the solvent viscosity of the tissue and matrix respectively. The  viscoelastic stress in the passive matrix has the form $f(\tau)\frac{(1-\phi)}{2}\boldsymbol\sigma^{'}$ where the scaling ensures that  contributions are non-vanishing only for the viscoelastic part ($\phi \approx -1)$, with $f(\tau)=(\gamma\tau)^2$  with $\tau$ the intrinsic time scale given by the ratio of shear viscosity and shear modulus (see SI.S7 for details).  The extra viscoelastic stress, $\boldsymbol{\sigma'}$, is assumed to evolve according to
\begin{align}
(\tau \partial_t 
+1){\boldsymbol\sigma'} -2 \eta_pD({\bf u})&= 0, 
\label{eqn:maxwell}
\end{align}
so that in the limit $\tau \rightarrow 0$, we recover the Newtonian fluid regime, whereas $\tau \rightarrow \infty$ leads to the Hookean solid regime. This allows us to easily recover the range of experimental conditions associated with the study~\cite{elosegui2023matrix} where $\tau$ is controlled by the mechanical properties of the passive matrix surrounding the growing spheroid. Finally, we impose the condition of incompressibility on the velocity $u$ so that
\begin{equation}
\nabla \cdot {\bf u} =0.
\end{equation}
The equations (1)-(4) suffice to track the evolution of the phase field description of the interface in terms of $\phi({\bf x},t)$, as well as the velocity ${\bf u} ({\bf x},t)$ and pressure $p({\bf x},t)$ and the matrix stress ${\boldsymbol\sigma'}({\bf x},t)$ given far field boundary conditions for these fields. We note that the phase-field description allows us track the interface by just following regions where $\nabla \phi$ is largest.

\paragraph*{Sharp-interface limit and linear stability analysis}
\label{sec:thin_interface_linear}
To characterize the morphological instability associated with invasion of the passive matrix by the active tissue, we first address the question of the linear instability of a growing straight interface. This calculation is analytically feasible in the sharp-interface limit, i.e.
$\epsilon/L\to 0$ of the phase-field system~(\ref{eqn:pfch}--\ref{eqn:maxwell}) (here $L$ corresponds to all other length scales in the problem, i.e. the width of the growing front, the size of the spheroid etc.) The resulting sharp-interface model takes the form of a Darcy-type tissue flowing into an incompressible Maxwell viscoelastic matrix, along with a capillary traction jump arising from the Korteweg stress. The full matched-asymptotic derivation and sharp--diffuse correspondence (see SI.S3), and allows for a linear stability calculation of a base interface that is steadily translating with  normal-modes transverse to the growing  state (as in section 4 of SI). In summary, in the sharp-interface limit we track the active-phase pressure and velocity $(p_a,\mathbf{v})$, the matrix pressure and velocity $(p_m,\mathbf{u})$, and the viscoelastic extra-stress $\boldsymbol{\sigma}'$.


 Crucially, the stability analysis is performed about a \emph{directed} base flow
$\mathbf{v}^{b}=v_0\,\hat{\mathbf{y}}$ (with $v_0>0$), which breaks rotational symmetry and distinguishes perturbations along the interface $\hat x$ and perpendicular to it, i.e. along $\hat y$. Therefore the most general \emph{linearized} Darcy law in the active tissue takes a tensorial form:
\begin{equation}
\nabla(\delta p_a)\;=\;-\mathbf{B}\,\delta\mathbf{v},
\qquad
\mathbf{B}=\mathrm{diag}(B_x,B_y),
\label{eq:darcy_tensor_main}
\end{equation}
where $\delta\mathbf{v}=\mathbf{v}-\mathbf{v}^{b}$ and $\delta p_a=p_a-p_a^{b}$ denote perturbations. In the sharp-interface model derived by linearizing the active Darcy forcing about $\mathbf{v}^{b}$ (see SI.S5), one has $B_x=\beta-\alpha/v_0$ and $B_y=\beta$, explicitly exhibiting the anisotropy induced by the base-state direction, noted before in a study that resembles ours at linear order, e.g. \cite{bogdan2018fingering}. The growth law in the tissue is
\begin{equation}
\nabla\cdot \boldsymbol{v} = \gamma \exp\!\big(-(\eta-y)/l\big), 
\label{eq:sharp_growth}
\end{equation} where $\eta$ is the interface, $l$ is a length scale where growth occurs, and $\gamma$ is the growth rate.
In the bulk matrix ($\phi\approx -1$), $\alpha(\phi)\to 0$ while the viscoelastic extra stress is present, and the outer limit reduces to an incompressible Maxwell medium:
\begin{equation}
\left\{
\begin{aligned}
\nabla\cdot \mathbf{u} &= 0,\\
\nabla\cdot\boldsymbol{\sigma} &= 0,\\
(\tau^{*}\partial_t+1)\boldsymbol{\sigma}'
&=\eta_p^{*}\big(\nabla\mathbf{u}+\nabla\mathbf{u}^{T}\big),
\end{aligned}
\right.
\qquad (\phi\approx -1).
\label{eq:maxwell_matrix_outer_main}
\end{equation}
where $\tau^{*}$ and $\eta_p^{*}$ are the dimensionless relaxation time and viscosity (see SI.S5). Here   $\boldsymbol{\sigma}:= -p_m\,\boldsymbol{I} +\eta_s^{*}\big(\nabla\bu+\nabla\bu^{T}\big) +\boldsymbol{\sigma}'$ where $p_m$ enforces mass conservation $\nabla \cdot {\bf u}=0$, $\eta_s^{*}$ is the (matrix) solvent viscosity, and $\boldsymbol{\sigma}'$ is the viscoelastic extra-stress evolving according to the Maxwell constitutive law (3). \\
In the sharp-interface limit, the Korteweg term converges to a capillary traction jump, and linearization about a flat interface yields the condition
\begin{equation}
\big[\![ {\bf n} \cdot \boldsymbol\sigma \cdot {\bf n}  ]\!] \;=\; -\,S\,\partial_{xx}\eta,
\label{eq:capillary_jump_linear}
\end{equation}
where $S$ is the surface tension. Together with the kinematic condition expressing continuity of normal velocity, which at linear order reads as
\begin{equation}
\partial_t \eta \;=\; v_y\big|_{y=\eta}\;=\;u_y\big|_{y=\eta},
\label{eq:kinematic_linear}
\end{equation}
 where $u_y,v_y$ are the transverse ($y$ component) velocity fields of the matrix and active tissue, respectively.

\textit{Dimensionless parameters:}
The physical problem is controlled by four key dimensionless parameters. i) A dimensionless growth length-scale: $L= \frac{l\,\beta \gamma}{\alpha}$, reflecting the ratio of the growth zone length and the length scale that arises from balancing the active stress and capillary forces, $\alpha/\beta \gamma$, ii) a scaled viscoelastic relaxation time-scale: $\tau^\ast = \tau\gamma$, characterizing the ratio of the passive relxation time of the matrix $\tau$ and the growth rate $\gamma$, iii) a scaled viscosity $\eta_s^\ast = \frac{\eta_s\,\beta\gamma^2} {\alpha^2}$ reflecting the balance between the viscous stresses and activity, iv) an effective Capillary Number: $\Ca = \frac{S\,\beta^2 \gamma^2}{\alpha^3}$ reflecting the balance between surface tension and active stresses. For typical experiments with carcinoids~\cite{elosegui2023matrix}: $L \in [0, 0.1]$ , $\tau^* \in [0.001,0.1]$, $\eta_s^\ast \in [10^{-2},10^{-7}]$, and $\Ca \in [0.1,0.5]$ (see SI.S5 for details).


\textit{Stability of a straight interface:}
To characterize the stability of the interface, we perturb it using the Fourier ansatz $\eta(x,t)=\eta_b(t)+\hat{\eta}\,e^{ikx+\Omega t}$ and use the same form for all fields in each of the phases with $\Omega$ being the complex growth rate, imposing decay away from the
interface.   Applying the kinematic condition, velocity continuity, tangential continuity, and the capillary traction jump yields the dispersion relation $\Omega(k;\Ca,\tau^*,L,\eta^*)=0$ (derived in SI.S5) written as:
\begin{equation}
-1 + \frac{1}{1+kL}+ kL+ \Omega+ k^3\Ca-\frac{2k^2\Omega\eta_s^{*}}{1+\tau^{*}\Omega}=0,
\label{eq:dispersion_relation_full_main}
\end{equation}
where $L= \frac{l \beta \gamma}{\alpha}$ is the dimensionless growth-localization length, where $l$ is the growth zone defined in (6), and $\eta_s^*= \eta_s\beta\gamma^2/\alpha^2$ is a dimensionless viscosity (scaled by the active stress). The interface is unstable when $\Re\,(\Omega(k))>0$.

\begin{figure}
    \includegraphics[width=1.0\columnwidth]{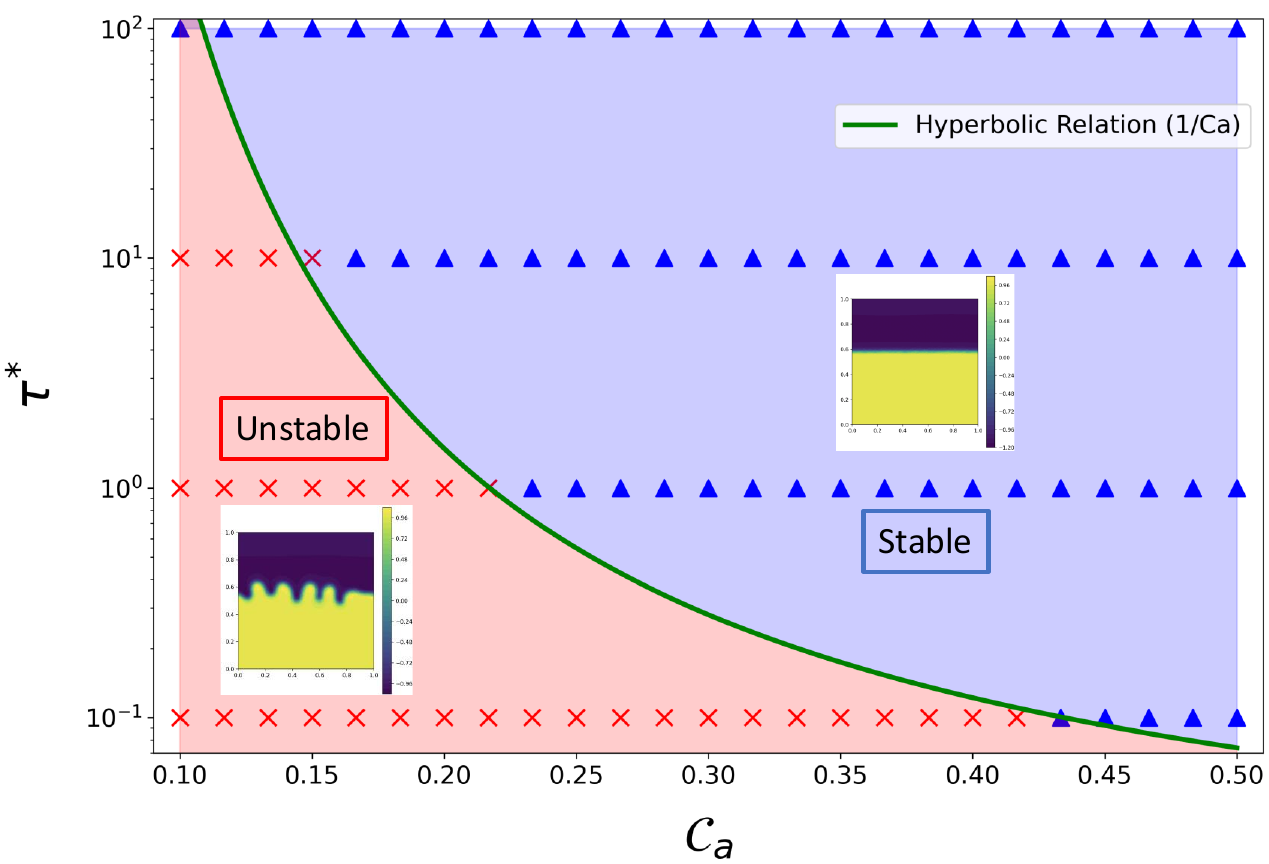}
    \caption{\textbf{Phase diagram obtained via predictions of the linear stability analysis and non-linear FEM simulations.}   Morphological phase diagram of the interface as a function of $\tau^*$ and $\Ca$. The blue portion of the phase diagram is the stable regime (i.e $\Re(\Omega(k))<0$) whereas the red portion is the unstable regime (i.e $\Re(\Omega(k))>0$), following (10); the curve that separates the two regime is given by a hyperbola (13). We then use FEM simulations to corroborate our results: the red crosses denote when the interface is unstable whereas the blue triangles correspond to the stable interface. The FEM simulations were based on solutions to (1-4). The ($\tau^{*}, \Ca$) used in the simulations for the stable and unstable snapshots are given by  $(100, 0.12)$ and $(0.1, 0.12)$ respectively. All the other parameter values are listed in the SI. }
    \label{fig2} 
\end{figure}

\begin{figure*}
    \includegraphics[width=2.2\columnwidth]{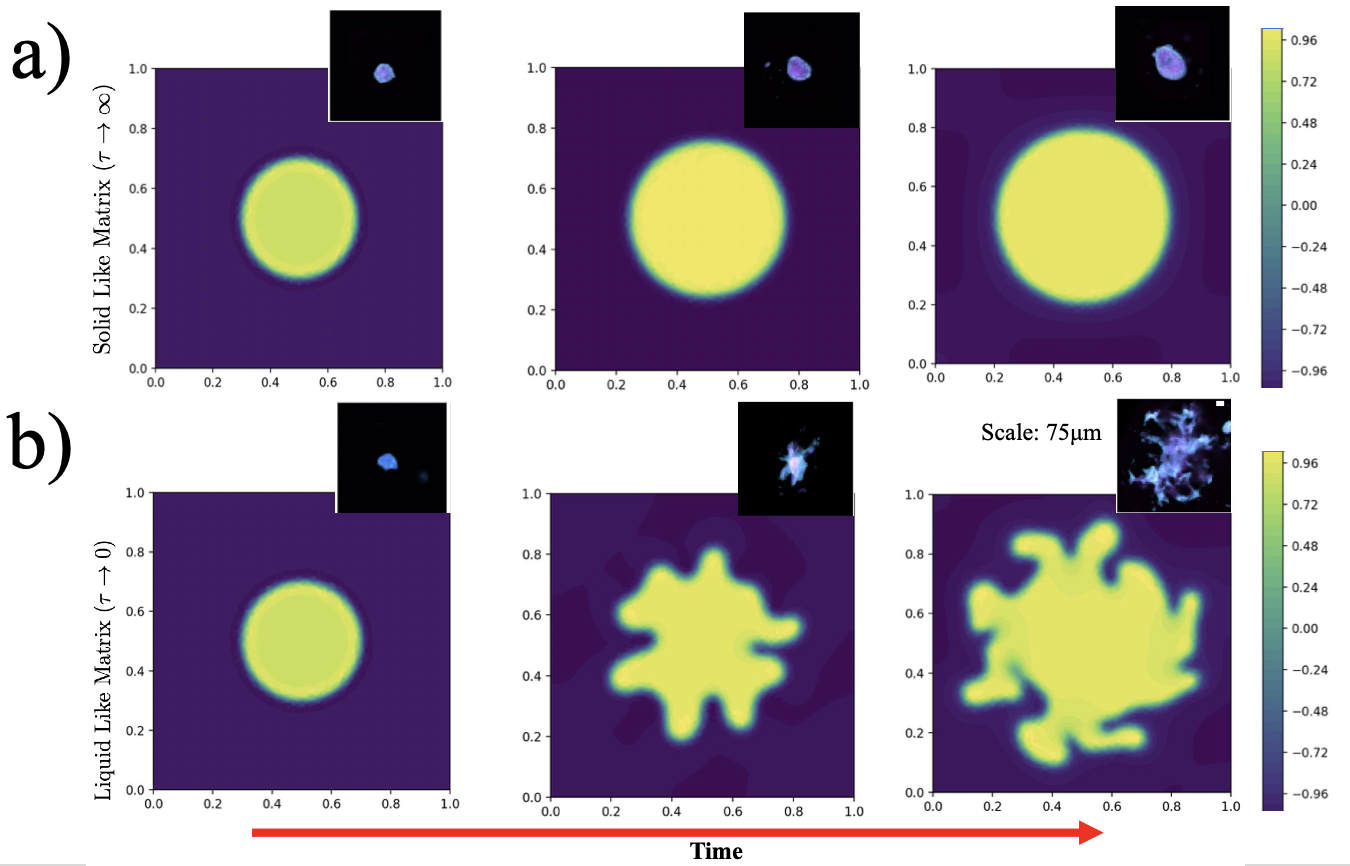}
    \caption{\textbf{Phase-field FEM simulations for the evolution of an invasive active-passive interface} (a) Two dimensional simulations of a growing tumor liquid spheroid pushing against an elastic matrix. The tissue grows without developing any instabilities consistent with the predictions of our theory (see the panel from left to right).  The ($\tau^{*}, \Ca$) used in the simulation is given by $(100, 0.12)$ (b) Same as (a) but with liquid-like matrix. The simulations are consistent with the theoretical predictions which predict the emergence of instabilites in a liquid-like matrix.  The inset within each figure shows the experimental micro-graphs from~\cite{elosegui2023matrix}. The ($\tau^{*}, \Ca$) used in the simulations are given by $(0.1, 0.12)$. The equations solved for both set of cases are (\ref{eqn:pfch}-\ref{eqn:maxwell}) and rest of the parameter values are listed in Table S1. }
    \label{fig3}
\end{figure*}

In the fast-relaxation / high-frequency regime $|\tau^{*}\Omega|\gg 1$, one may approximate
$(1+\tau^{*}\Omega)^{-1}\simeq (\tau^{*}\Omega)^{-1}$, and~\eqref{eq:dispersion_relation_full_main} reduces to
\begin{equation}
\Re\,(\Omega(k))\;\simeq\;
1- kL - \frac{1}{1+kL} + \frac{2k^2\eta_s^\ast}{\tau^\ast} - k^3 \Ca.
\label{eq:dispersion_relation_asympt_main}
\end{equation}
As $k\to 0$, $\Re\,(\Omega(k))\to 0$, corresponding to a neutrally stable translation mode of the base state, while as $k\to \infty$ the $-k^3\Ca$ term stabilizes short wavelength perturbations. To make the short-wave cutoff explicit, note that at large $k$ the dominant balance  is between the capillary term and the
viscoelastic contribution, so that
\begin{equation}
\Re\,(\Omega(k))\sim -\,k^3\Ca + \frac{2k^2\eta_s^{*}}{\tau^{*}},
\qquad (k\to\infty,\ |\tau^{*}\Omega|\gg 1).
\end{equation}
Thus, unstable modes must satisfy (see SI.S6 for details)
\begin{equation}
k \;\lesssim\; k_c
\;:=\;
\frac{2\eta_s^{*}}{\tau^{*}\Ca}.
\label{eq:kc_main}
\end{equation}
Equation~\eqref{eq:kc_main} summarizes the key hyperbolic trend: increasing matrix relaxation time $\tau^{*}$ (more elastic-like response) reduces the unstable band,
while decreasing $\tau^{*}$ (more liquid-like response) broadens it, consistent with the experimental observations in~\cite{elosegui2023matrix}. The full
stability criterion (including the non-uniformity of the small-$\tau^{*}$ limit and the re-stabilization at $k\gg k_c$) is given SI.S6.

\textit{Phase-field Simulations:} To corroborate the linear stability analysis in the sharp-interface limit, and go beyond it by following the dynamics of interfacial instability, we now turn to phase field simulations of the equations [\ref{eqn:pfch}-\ref{eqn:maxwell}]. We considered two geometries for the phase-field simulations. Firstly, we calibrate the simulations with the analytical predictions of linear stability analysis, by solving Eqns. [\ref{eqn:pfch}-\ref{eqn:maxwell}] in a square box with periodic boundary conditions in $x$ and zero flux conditions in $y$. The initial $\phi$ field was initialized by $\phi=1$ for $y< \frac{L}{2}$ and $\phi=-1$ for $y> \frac{L}{2}$. Consistent with the theory, the active drop remains stable in an elastic regime (i.e high $\tau^{*}$) and is unstable in the liquid regime (i.e low $\tau^{*}$). FIG.\ref{fig2}b shows the ($\tau^{*}, \Ca$) phase diagram, where for higher $(\tau^{*}, \Ca)$, the interface is stable and vice-versa. 

In the second case, we initialized the active phase, $\phi=1$, on a circular domain of radius $\frac{L}{4}$ with periodic boundary conditions in both $x$ and $y$ (see SI.S7), to mimic the experimental conditions by simulating the condition where the initial active droplet is a circular domain embedded in a passive matrix, with zero  velocity on all four boundaries. The simulations are in good agreement with the experimental observations \cite{elosegui2023matrix}: for a solid carcinoid embedded in a solid-like matrix, no interfacial instabilities arise, while when the carcinoid is embedded in a liquid-like matrix, the droplet develops instabilities at the growing interface (See Fig. \ref{fig3}). We note that the characteristic finger width is controlled by the interfacial scale $\epsilon$. \\

\noindent \textit{Discussion:} Utilizing continuum mechanics and FEM simulations, we have provided a theoretical basis for the experimental observation of how the viscoelastic matrix regulates the interfacial properties of an invasive growing tumor spheroids. The \textit{in-vitro} experiments showed that when the elasticity of the matrix is enhanced (reduced), the interface becomes morphologically stable (unstable) \cite{elosegui2023matrix}. Our continuum model takes the form of a theory for a growing two-dimensional active Darcy-like droplet pushing against a passive Maxwell-like viscoelastic matrix. The linear stability analysis of the dynamical field equations for small perturbations reveals that short-wavelength instabilities are always stable because of surface tension regardless of the matrix properties. However, long-wavelength instabilities are regulated by the stress relaxation time scale, $\tau$, of the matrix with the interface being morphologically stable (unstable) if the matrix is elastic (liquid-like). To track the non-linear evolution of the interface, we developed a custom phase-field FEM simulation of an active droplet invading a viscoelastic matrix. The simulations are in accordance with predictions of the theory and are consistent with the experimental observations. Taken together, our study provides a physical basis for the interfacial control of far-from equilibrium active systems.    \\

\textbf{Acknowledgements:} SS acknowledges the funding support from Dana Farber Cancer Institute (Prof. C.Z. Zhang), and LM thanks the Simons Foundation and the Henri Seydoux Fund for partial financial support.

\textbf{Code Availability:} The custom FEniCS code has been hosted on GitHub at the following LINK.



\bibliography{undulation}
\bibliographystyle{apsrev4-2}

\end{document}